\documentclass[prd,twocolumn]{revtex4}
\usepackage{graphicx, epsfig}
\usepackage{color}
\usepackage{mathrsfs}
\usepackage{bm}
\usepackage{amsmath,amssymb}
\usepackage{tikz}
\usepackage{subfig}

\newcommand{\be}{\begin{equation}}
\newcommand{\ee}{\end{equation}}
\newcommand{\ba}{\begin{eqnarray}}
\newcommand{\ea}{\end{eqnarray}}

\newcommand{\gapp}{\mathrel{\raise.3ex\hbox{$>$}\mkern-14mu
              \lower0.6ex\hbox{$\sim$}}}
\newcommand{\lapp}{\mathrel{\raise.3ex\hbox{$<$}\mkern-14mu
              \lower0.6ex\hbox{$\sim$}}}

\begin{document}
\title{String Production in the Abelian Higgs Vacuum}
\author{Ayush Saurabh$^\dag$, Tanmay Vachaspati$^{\dag *}$}
\affiliation{
$^\dag$Physics Department, Arizona State University, Tempe, AZ 85287, USA. \\
}

\begin{abstract}
\noindent
We numerically study string production by evolving classical Abelian-Higgs gauge 
	field wavepackets. Initial conditions are constructed for the propagation of a
	single wavepacket and for the collision of two 
	wavepackets. We identify regions of parameter space that lead to prompt 
	production of strings from a single wavepacket. The collision of two sub-critical
	wavepackets can also lead to the production of strings in certain regions of
	parameter space.
\end{abstract}
\maketitle


Topological defects such as kinks, strings and magnetic monopoles are classical
solutions in a wide range of field theories. In quantum theory, topological defects can
be viewed as a bound state of a large number of quanta. The interpretation of solitons as 
particles is most explicitly known in the sine-Gordon
model. In that case, the operators that create and destroy solitons (which are fermions), 
can be written in terms of particle quanta that are bosons. The question of interest in this 
paper is if it is possible to assemble particles to make strings? And if so, can we say something 
about the initial conditions necessary to produce strings?

The transition from particles to solitons is difficult to treat because particles are described by
quantum field theory whereas solitons are described by classical field theory. However, from
a practical standpoint, we often produce high occupation number states of quantum 
particles that behave quasi-classically. For example, by sending currents into a
light bulb we produce light that can be described as classical radiation using
Maxwell's equations. Thus it is relevant to consider the production of solitons in
the scattering of classical waves or wavepackets. We will restrict our attention to
this situation and ask what {\it classical} initial conditions lead to the production of
solitons in the final stage.

These questions were addressed in Ref.~\cite{TanmayCreation} 
for an SO(3) field theory, where
incoming wavepackets led to the production of magnetic monopoles. While the
possible production of magnetic monopoles is exciting, it is in the realm of speculative
physics because we don't know if Grand Unified Theories are correct. On the other
hand, strings are closer to reality since we do have superconductors in which (gauge) 
strings exist. In this paper we focus on the production
of gauge U(1) strings, where the class of initial conditions we use are 
motivated by the initial conditions of Ref.~\cite{TanmayCreation}.

There are several aspects of the string creation problem that differ from the monopole
creation problem. In the latter, once monopole-antimonopole pairs are created with
enough energy, they fly apart and survive indefinitely. On the other hand, only closed
loops of string
can be created. These oscillate, radiate, collapse, and survive only for a finite
amount of time. If some of the loops are produced with large angular momentum, they
live for longer but eventually decay. A second difference is that the properties of the
string network that is produced change with time because the strings interact with
each other and intercommute to form smaller loops. 


We introduce the field theory and string solution in Sec.~\ref{model} followed
by our choice of initial conditions in Sec.~\ref{initial}. The computational methods
used in our analysis are described in Sec.~\ref{computational} and then we present
our results in Sec.~\ref{results}. We conclude in Sec.~\ref{conclusions}.

\section{Abelian Higgs Model and Strings}
\label{model}


We consider the Abelian Higgs model given by the Lagrangian,
\begin{equation}
	\mathcal{L}=-\frac{1}{4} F_{\mu \nu} F^{\mu \nu} + 
	\frac{1}{2}| D_\mu\phi |^2-\frac{\lambda}{4} \left( |\phi |^2-\eta^{2} 
	\right)^{2} 
\end{equation}
where $\phi = \phi_1 + i \phi_2$ is a complex scalar field, 
$D_\mu = \partial_\mu + i e A_\mu$,
$A_\mu$ is the gauge field with field strength tensor 
$F_{\mu \nu} = \partial_\mu A_\nu - \partial_\nu A_\mu$, and 
$\lambda$ and $e$ are coupling constants.  
The energy density in the fields is,
\be
	E = \frac{1}{2}|D_0 \phi |^2+ \frac{1}{2}|D_i \phi |^2
	+  \frac{1}{2} ({\bf E}^2+{\bf B}^2) +
	\frac{\lambda}{4} \left( |\phi |^2-\eta^{2} \right)^{2}
\ee
where $E_i = F_{0i}$ is the electric field and $B_i = \epsilon_{ijk}F_{jk}/2$
is the magnetic field.

Topological string solutions in the Abelian Higgs model are well known. The
solution for a straight string along the $z-$axis is,
\be
\phi = \eta f(r) e^{i\theta}, \ \ 
A_i = v(r) \epsilon_{ij} \frac{x^j}{r^2} \ \ (i,j=1,2)
\ee
where we work in cylindrical coordinates $r=\sqrt{x^2+y^2}$,
$\theta=\tan^{-1}(y/x)$, $f(r)$ and $v(r)$ are profile functions that vanish
at the origin and go to 1 asymptotically.
The energy per unit length (also the tension) of the string is given by
\be
\mu = \pi \eta^2 F(\beta )
\ee
where $\beta \equiv 2\lambda/ e^2$. The function $F(\beta)$ is known
numerically and is a smooth, slowly varying function. We also have 
$F(1)=1$ in the so-called BPS limit when the scalar mass in the model,
$m_S = \sqrt{2\lambda} \eta$ equals the vector mass,  $m_V =  e \eta$.
For $\beta$ not too large, the thickness of the scalar fields in the string is 
$\sim m_S^{-1}$ and of the vector fields is $\sim m_V^{-1}$.


The string is characterized by a topological winding number that is defined by
%
\be
n = \frac{-i}{2 \pi \eta^2} \oint dx^i \phi^* \partial_i \phi = 
\frac{1}{2 \pi} \oint \frac{d \theta}{dl} dl  
\label{windingdefn}
\ee
where $\theta$ is the phase of the scalar field at a given point on the 
contour and $l$ denotes the parameter along the integration curve.
%


%

\section{Initial Conditions}
\label{initial}


We base the initial conditions for our simulations on those used for 
monopole-antimonopole production \cite{TanmayCreation}. 
%
%
We adopt the temporal gauge for all our simulations, that is $A_0=0$,
and construct circularly polarized gauge wavepacket configurations 
(not solutions) that propagate along the $\pm z-$axis. 
Consider the form of a wavepacket propagating in the $-z$ direction,
\begin{eqnarray}
 {\cal A}_x &=& \partial_y f_1 (\omega  f_2  - \partial_z f_2 ) \cos( 
	\omega (t+z- z_0)) \nonumber \\
 {\cal A}_y &=& \partial_x f_1 (\omega f_2  +\partial_z f_2 ) \sin(\omega 
	(t+z-z_0)) \nonumber \\
 {\cal A}_z &=& \partial_x\partial_y f_1 f_2  [\cos(\omega (t+z-z_0)) 
 -\sin(\omega (t+z-z_0 ))] \nonumber
\end{eqnarray}
where $f_1=f_1(x,y)$, $f_2=f_2(t+z-z_0)$ will be specified below, and $z_0$ 
determines the initial ($t=0$)
location of the wavepacket along the $z-$axis. Now the initial conditions for 
the gauge fields and their time derivatives are,
\ba
A_i (t=0,{\bf x}) &=& {\cal A}_i (t=0,{\bf x}), \\
\partial_t A_i (t=0,{\bf x}) &=& \left [ \partial_t {\cal A}_i (t,{\bf x}) \right ]_{t=0}
\ea
This form for the gauge fields satisfies $\nabla\cdot {\bm A} =0$ which will be
useful later when we discuss Gauss constraints.



We can also construct a wavepacket traveling in the $+z$ direction in 
a similar manner. To do this,  we write the formulae
in terms of $ f_3  (t-(z+z_0))$:
\begin{eqnarray}
 {\cal A}'_x &=& \partial_y f_1 (-\omega '  f_3 - \partial_z f_3 )\cos (	
	\omega' (t-z-z_0) \nonumber  \\
 {\cal A}'_y &=& - \partial_x f_1 (\omega ' f_3 -\partial_z f_3 )\sin( 
	\omega '(t-z-z_0)) \nonumber \\
 {\cal A}'_z &=& \partial_x\partial_y f_1  f_3  (\cos(\omega' (t-z-z_0)) 
	-\sin(\omega ' (t-z-z_0)) \nonumber
\end{eqnarray}
And these can be used to construct initial conditions for a wavepacket
that propagates in the $+z$ direction as above.


We choose profile functions in a manner that localizes the gauge 
wavepacket in all directions;
\begin{equation}
f_1(x,y) = a \, \exp \left [ -\frac{x^2+y^2}{2w^2} \right ]
\end{equation}
\begin{equation}
f_2(t+z-z_0) = \exp \left [ - \frac{(t+z-z_0)^2}{2w^2} \right ]
\end{equation}
\begin{equation}
f_3(t-z-z_0)) = \exp \left [ -\frac{(t-z-z_0)^2}{2w^2} \right ]
\end{equation}
where $a$ is the amplitude and $w$ is the width of the wavepacket.

The initial conditions for the scalar field are ``trivial'',
\be
\phi (t=0,{\bf x}) = \eta, \ \ [\partial_t \phi (t,{\bf x}) ]_{t=0} = 0.
\ee

The free parameters in the initial conditions are $z_0$, $a$, $w$, 
$\omega$
and $\omega'$. For our simulations, we will rescale these parameters as 
follows:
\begin{equation}
	z_0 = \frac{ \bar{z}_0}{\eta},\, a = \frac{ \bar{a}}{\eta},\, w = 
	\frac{ \bar{w}}{\eta},\, \omega = { \bar{\omega}}{\eta},
	\, \omega^\prime = { \bar{\omega}^\prime}{\eta}
\end{equation}
The dimensionless parameters $\bar{z}_0$, $\bar{a}$, $\bar{w}$, 
$\bar{\omega}$, and $\bar{\omega}^\prime$ above are varied in our code.  
In addition, the Abelian Higgs model has the parameters
$e$, $\lambda$ and $\eta$. However, by field and coordinate rescalings,
there is only one model parameter given by the ratio of scalar and vector
masses, $\beta = m_S^2/m_V^2 = 2\lambda / e^2$. 


\section{Computational Techniques}
\label{computational}

Following the numerical relativity based approach developed in 
\cite{TanmayCreation}, we introduce a new dynamical variable 
$\Gamma = \partial_i A_i$. Then the field variables are:
$\phi$, $A_i$ and $\Gamma$, altogether 6 functions.
The equations of motion for these variables are,
\begin{align}
	\partial_t^2 \phi_a &= \nabla^2 \phi_a
	- e^2 A_i A_i \phi_a - 2e \epsilon_{ab}\partial_i \phi_b A_i - e 
	\epsilon_{ab} \phi_b \Gamma \nonumber \\
	&	- \lambda (\phi_b\phi_b- \eta^2)\phi_a \\
	\partial_t F_{0i} &= \nabla^2 A_i - \partial_i \Gamma +  e 
	(\epsilon_{ab} \phi_a \partial_i \phi_b + e A_i \phi_a \phi_a) \\ 
	\partial_t \Gamma & = \partial_i F_{0i} - g_p^2 [ \partial_i 
	F_{0i}  + e \epsilon_{ab} \phi_a \partial_t \phi_b ]
\label{Gammaeq}
\end{align}
where $a=1,2$, $\epsilon_{ab}$ is the Levi-Civita tensor with 
$\epsilon_{12} = 1$, $F_{0i} = \partial_t A_i$ in the temporal gauge, and 
$g_p^2$ is a new parameter introduced for numerical stability.  The idea is
that the square bracket in Eq.~(\ref{Gammaeq})
vanishes in the continuum because of the Gauss constraints. However,
it may not vanish upon discretization. By writing the equations in the above
form with the auxiliary function $\Gamma$, we obtain improved numerical
stability~\cite{NumericalRelativity}. The value of the parameter $g_p^2$ is
chosen by numerical experimentation; we have set $g_p^2=0.75$ in our 
simulations. The initial conditions for the auxiliary function $\Gamma$ 
follow
from the choice of initial conditions for the gauge field,
\be
\Gamma (t=0,{\bf x})=0.
\ee

For our analysis, we discretized these equations on a $256^3$ lattice with 
lattice spacing $\Delta x=0.05$ and time step size $\Delta t=\Delta x/4$.  
The difference equations were solved using
the explicit Crank-Nicholson method with two iterations.
To reduce computation times, we parallelized our numerical code. 
As a check of our evolution code, we find that the total energy inside the 
box is conserved to within 1\% during the entire evolution period.

In addition to the evolution of equations, we developed a tracking code which 
detects strings and calculates the number of loops that are present in the simulation 
domain at any given time. The program calculates the phase winding as defined
in Eq.~(\ref{windingdefn}) on every plaquette of the lattice. A non-zero winding
on a plaquette implies that a string passes through the plaquette and enters/exits 
the corresponding cells. The program then connects the strings and records the 
properties of the loops.

The string tracking algorithm is the same as used in earlier work
~\cite{VilenkinVachaspati, TanmayLevonGeodesic} but with one subtlety.
In calculating the winding as in Eq.~(\ref{windingdefn}),
we have to find the discretized value of $d\theta$ along the links of the
lattice. Generally one uses the ``geodesic rule'' and the phase difference
between lattice sites $i$ and $i+1$ is 
\be
d\theta \to \Delta\theta \equiv \theta_{i+1}-\theta_i + 2\pi k
\ee
where $k=0,\pm 1$ is chosen to minimize $|\Delta\theta |$. However, this rule
ignores the case when $| \Delta \theta | = \pi$. The justification in earlier works
has been that this possibility is of zero measure. In our case, however,
this situation arises quite frequently. The reason can be seen from the
equations of motion and the initial conditions. We start out with $\phi_1=\eta$
and $\phi_2=0$ {\it i.e.} $\theta=0$ throughout the lattice. The equations of
motion are such that they tend to preserve $\phi_2=0$,
and all the non-trivial dynamics is in the $\phi_1$ variable, at least at
early times. Now $\phi_1$ can become negative. When $\phi_1$ differs
in sign at neighboring lattice sites, this gives a phase difference of exactly $\pm \pi$
and the geodesic rule is ambiguous. 
In evaluating the winding number, we choose $+\pi$ or $-\pi$ with equal probability.

\section{Results}
\label{results}

As we have discussed above, the problem contains 1 model parameter,
namely $\beta$, and 5 initial condition parameters. 
We will fix some of
these parameters and scan over a range of a few parameters. We set
\be
\bar{z_0}=1.8, \ \ \bar{w}=0.6, \ \ \bar{\omega}^\prime = \bar{\omega}.
\ee
We have explored,
\be
\beta \in [0.08,8.0], \ \ \bar{a} \in [0.6,7.0], \ \ \bar{\omega} \in 
[0.2,8.0].
\ee
We did not see any qualitative changes as we varied $\beta$ 
(see below) and so for most of our runs we set $\beta=1$,
equivalently $e=0.5$, $\lambda=0.125$. We also chose $\eta=1$  and this 
sets the length scale in the simulation. 




From our initial runs, we found that energy is condensed into strings even 
from a single wavepacket, {\it i.e.} without scattering two gauge wavepackets.
We will call this ``prompt string production'' and it
is reminiscent of the discovery in~\cite{Hindmarsh:2000kd} 
that strings may be formed due to purely gauge field fluctuations during a 
phase transition. In the next subsection, we will explore prompt string
production and find that there are regions of parameter space where
prompt production does not occur. We will then move on to explore
this region of parameter space and find a sub-region where strings
are produced when wavepackets collide.


\begin{figure*}[ht]
\includegraphics[height=0.24\textwidth,angle=0]{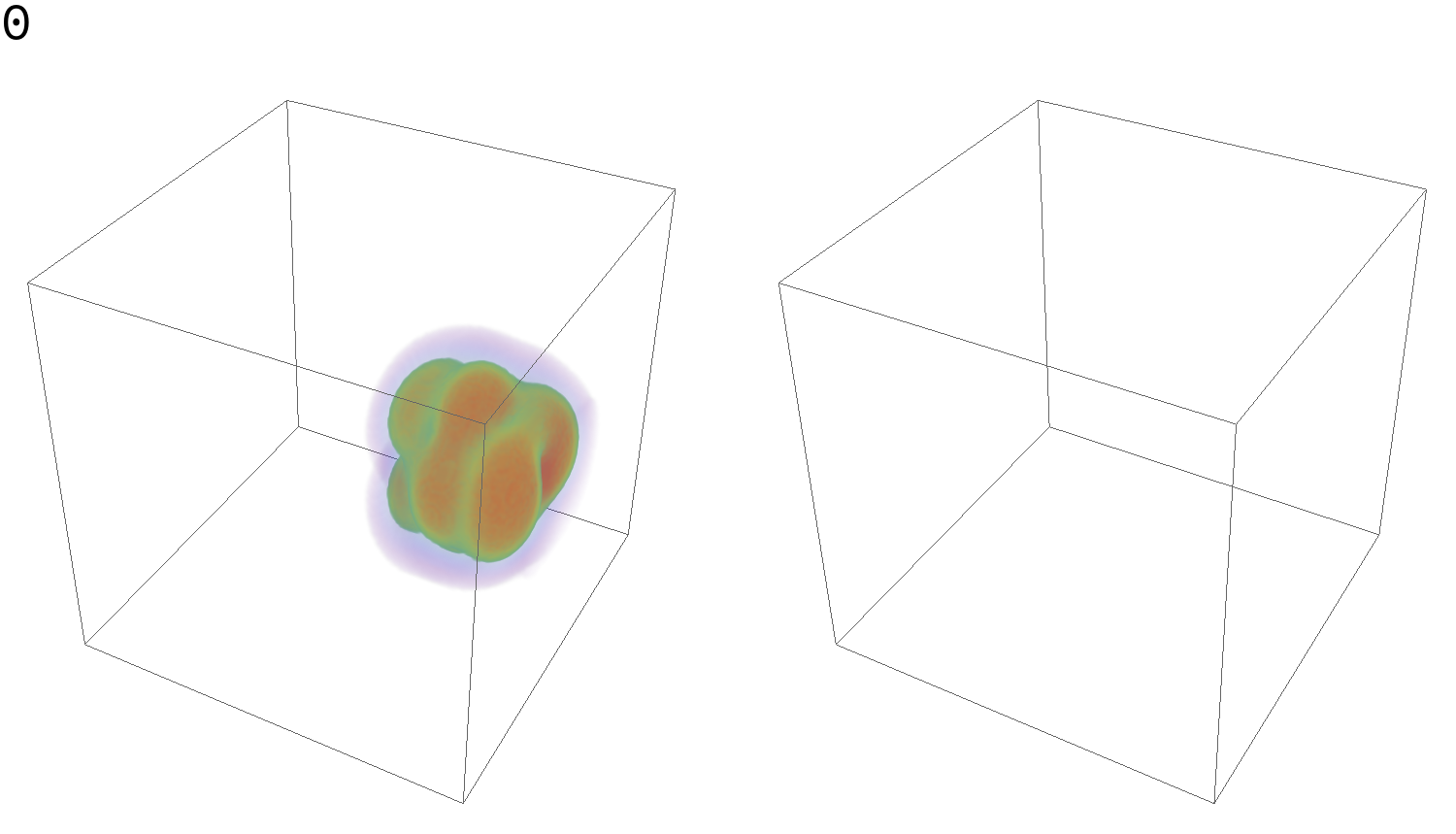}
\includegraphics[height=0.24\textwidth,angle=0]{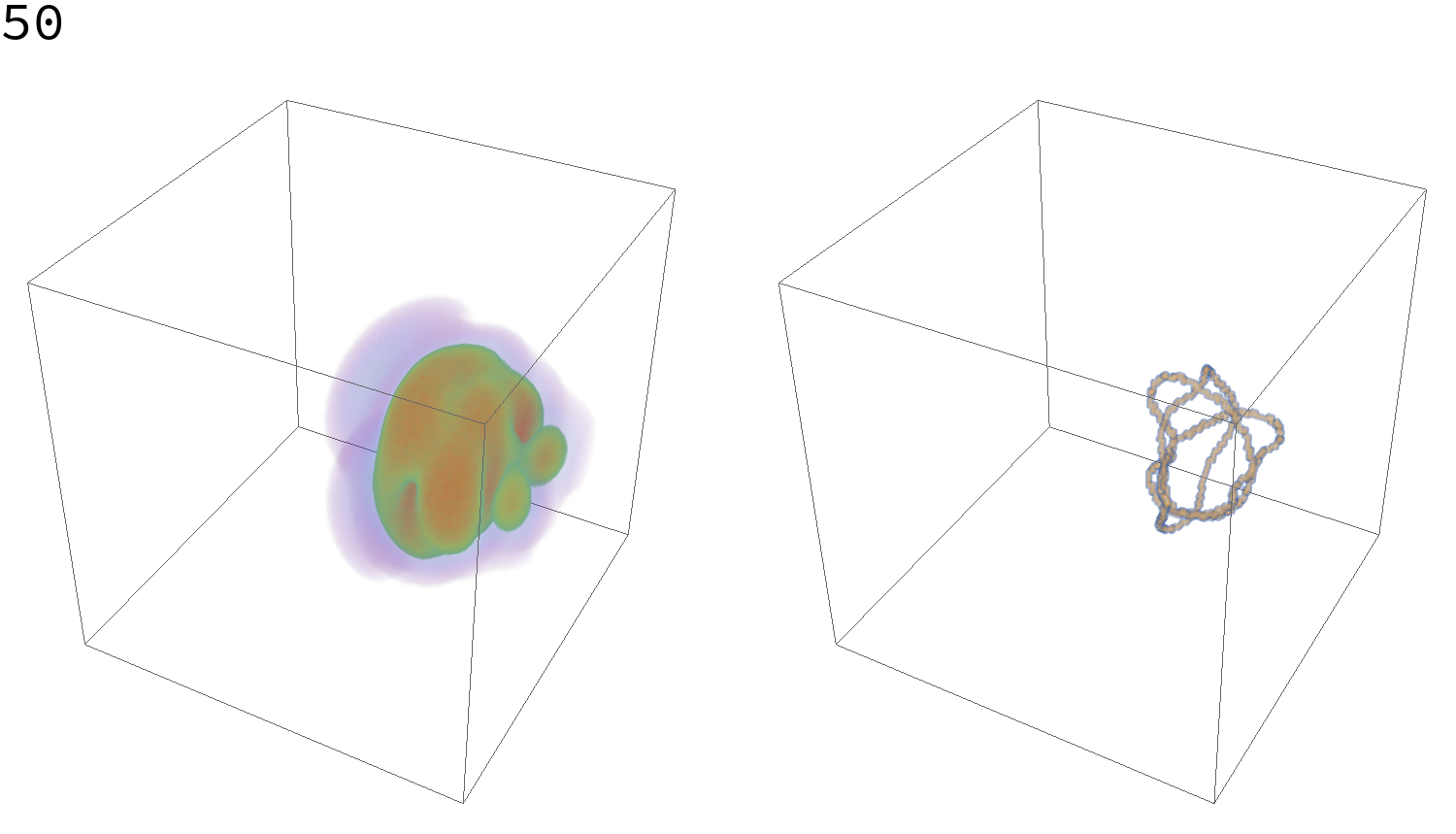}\\
\includegraphics[height=0.24\textwidth,angle=0]{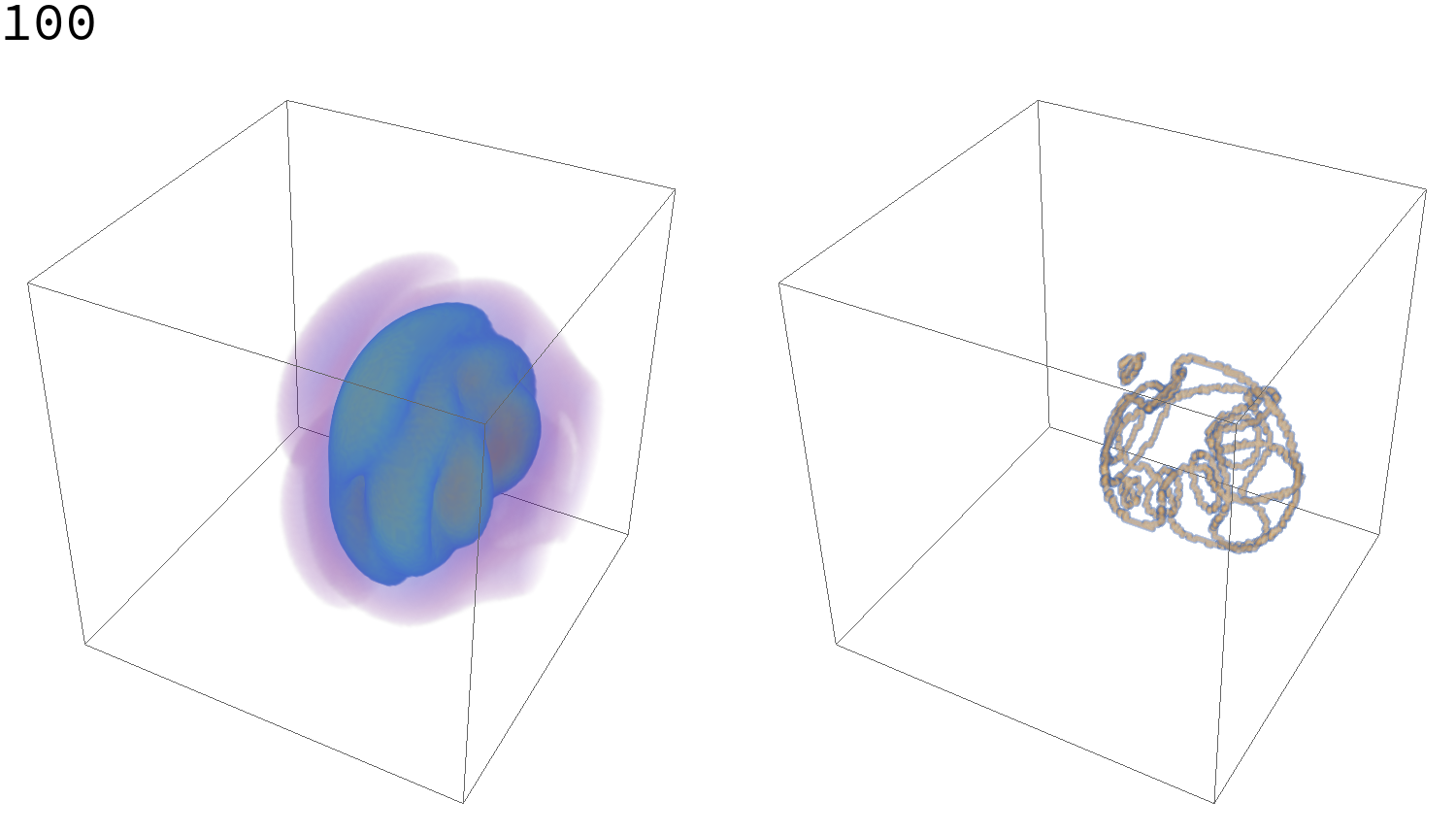}
\includegraphics[height=0.24\textwidth,angle=0]{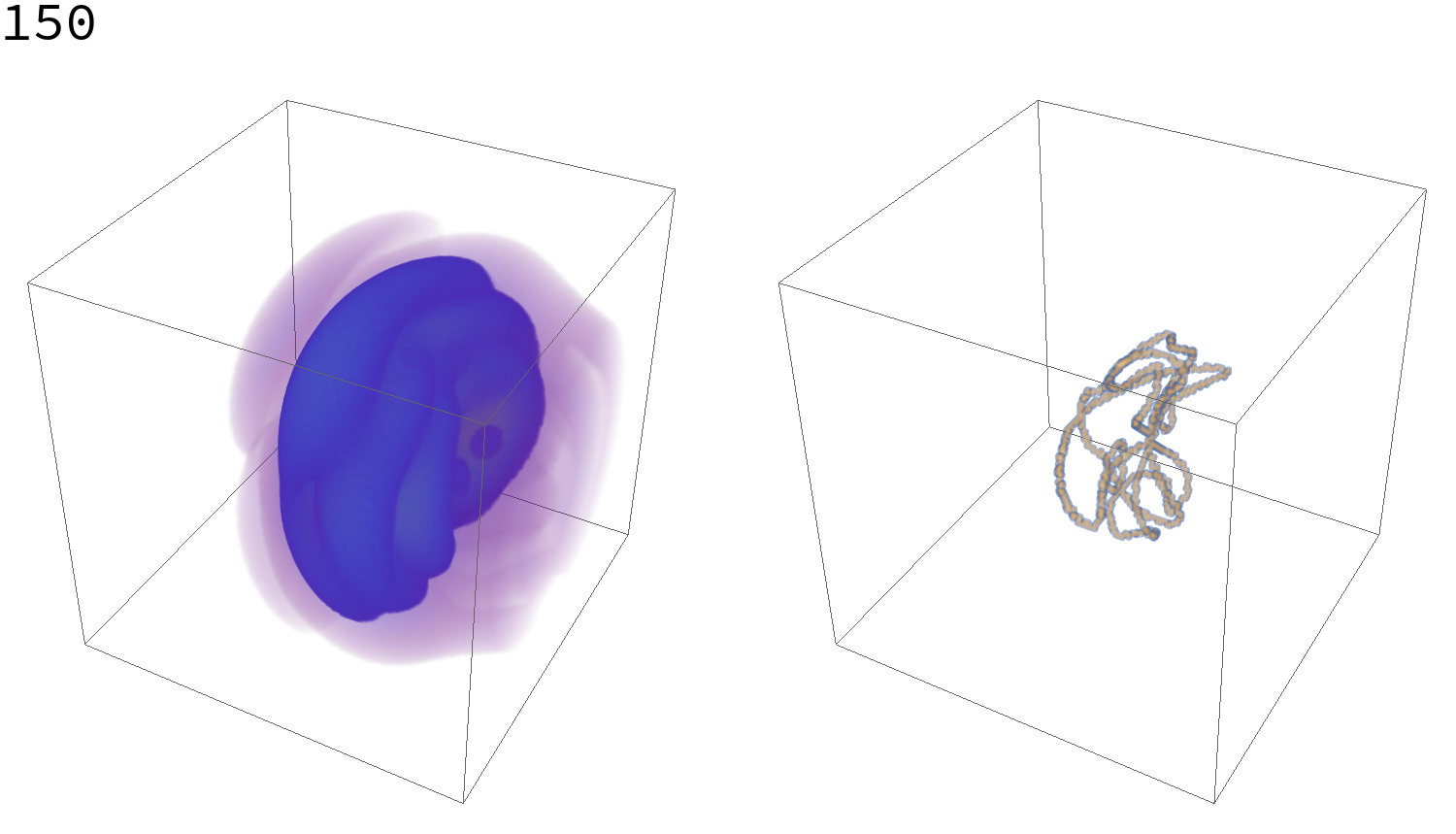}\\
\includegraphics[height=0.24\textwidth,angle=0]{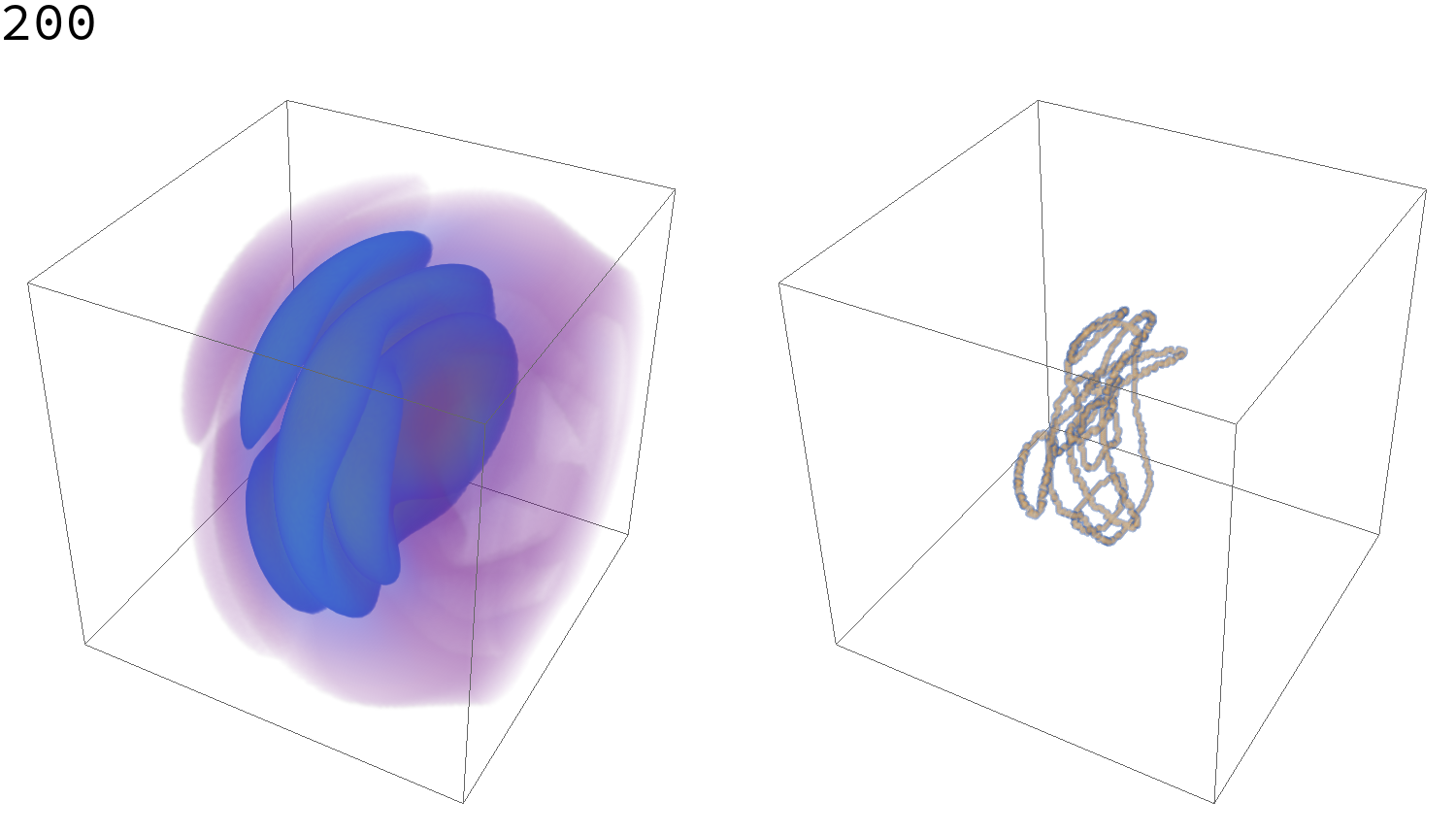}
\includegraphics[height=0.24\textwidth,angle=0]{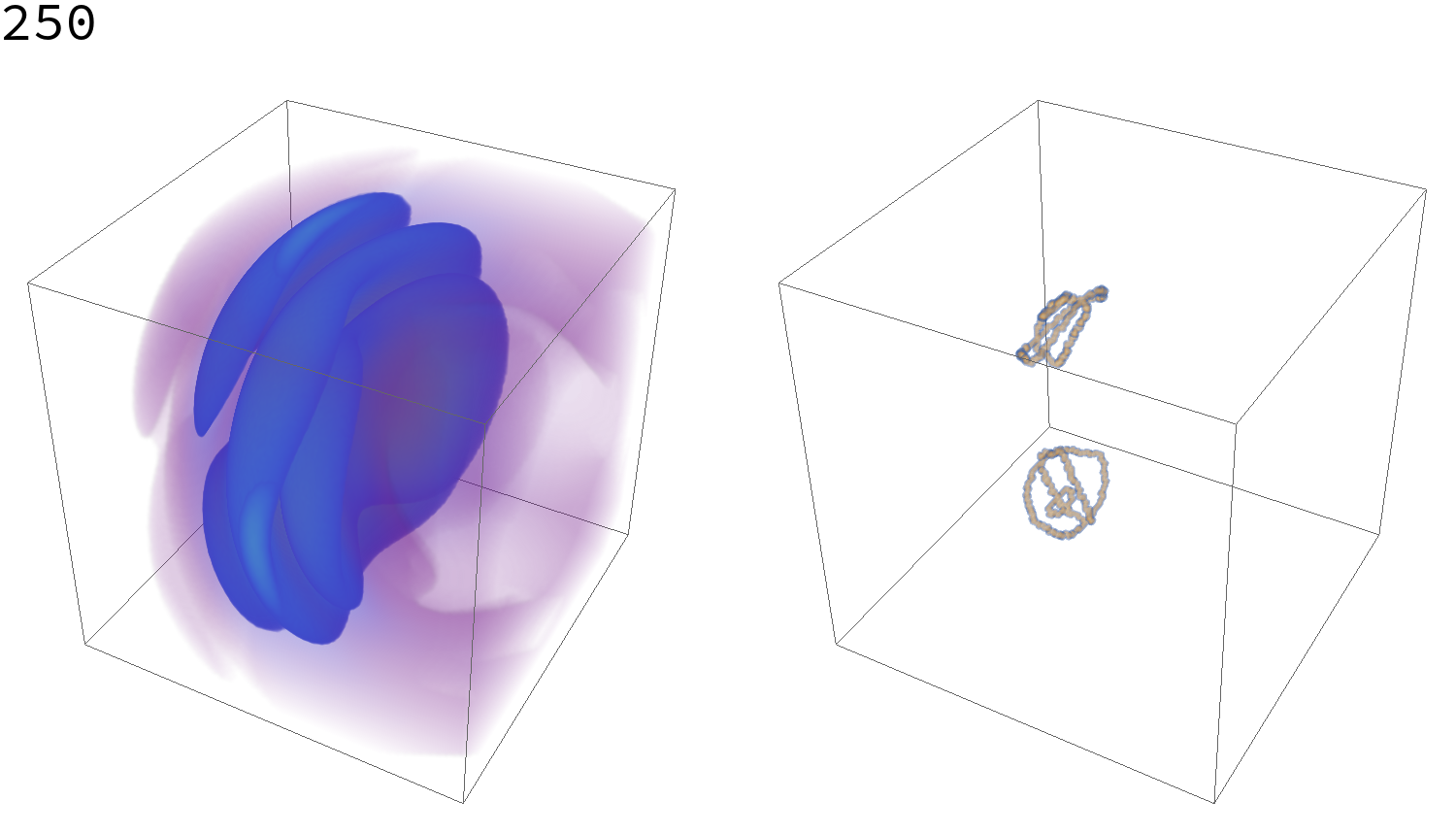}
\includegraphics[height=0.07\textwidth,angle=0]{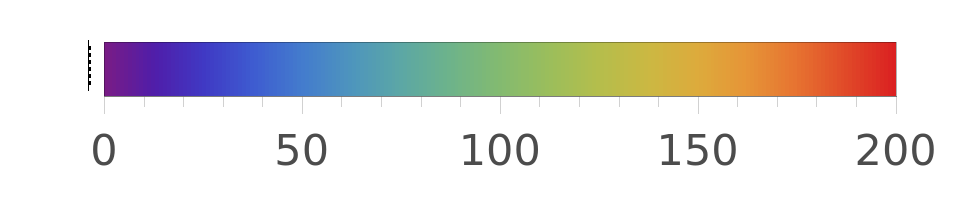}
\caption{Total energy density (boxes on the left) and 
winding (boxes on the right) at different time steps for the case of one 
	pulse for $\bar{a}=6.215$, $\bar{\omega}=2.0$, $ \lambda = 0.125$, 
	and $ E_{\rm Single} = 4000$. The $116^3$ boxes shown here are smaller 
	than the full lattice ($256^3$).}
\label{SinglePulseMovie}
\end{figure*}

\subsection{Prompt string production}
\label{prompt}

For the single pulse case, after fixing the parameters of the theory, we 
managed to analytically find expression for total energy in the simulation 
domain for the wavepacket profiles shown in the previous section. It is as 
follows
\be
E_{\rm Single} = \frac{\bar{a}^2 \pi^{\frac{3}{2}} \eta \, (9+10 \bar{w}^2 
\bar{\omega}^2 + 4 \bar{w}^4 \bar{\omega}^4 + 2 e^2 (\bar{w}^2+\bar{w}^4 
\bar{
	\omega}^2) )}{8 \bar{w}^3}
\label{EnergySingle}
\ee
where $\gamma \equiv w\omega$.
With this expression, we can trade one of the parameters for the total
energy.



In Fig.~\ref{SinglePulseMovie} we show the prompt production of strings at various times
during the evolution. In the first frame, there is energy density of the wavepacket but no
strings. Some time steps later, the scalar field has adjusted to the gauge wavepacket and
strings, as detected by topological winding, are produced. As the system evolves further,
the dense network of strings chops itself up and decays.

\begin{figure}[h]
\includegraphics[width=0.45\textwidth,angle=0]{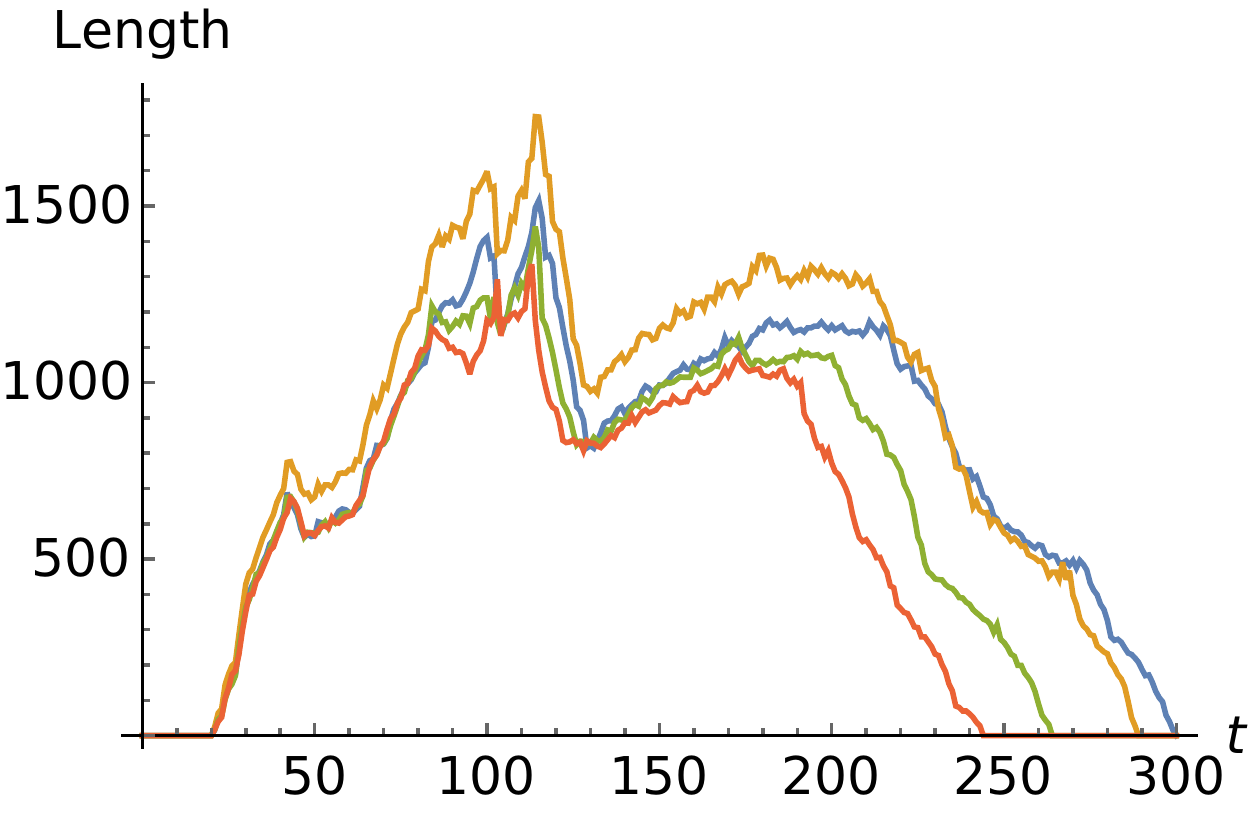}
\caption{
	Total length of strings (in units of number of lattice points) as 
	a function of time(-steps) for $\lambda=0.01$ (blue), 
	$\lambda=0.125$ (orange), $\lambda=0.50$ (green), and 
	$\lambda=1.0$ (red). All the other kinematic parameters are kept 
	fixed with $\bar{a} = 6.215$ and $\bar{\omega} = 2.0$.
}
 \label{LengthLambda}
\end{figure}

\begin{figure}[h]
  \includegraphics[height=0.30\textwidth,angle=0]{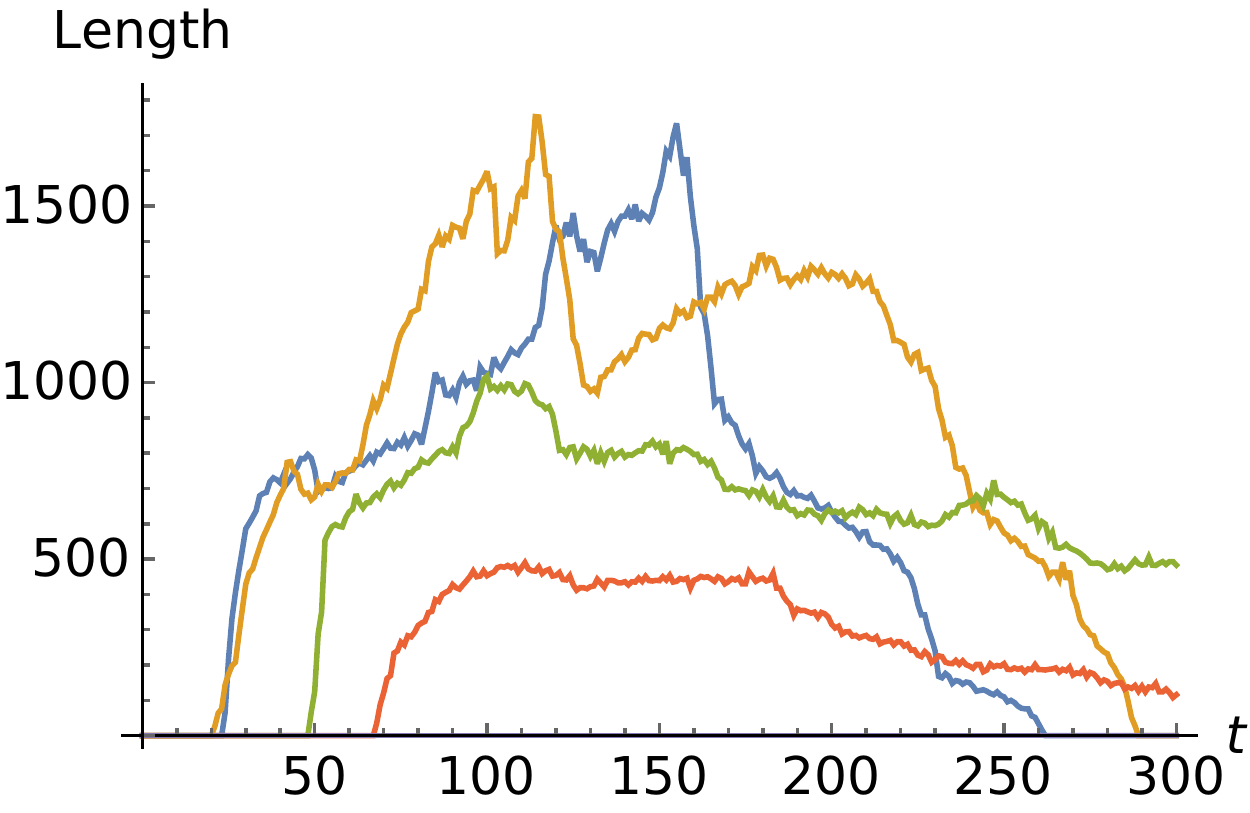}
  \caption{Total length of strings (in units of number of lattice points) 
	as a function of time(-steps) for $\bar{\omega}=0.1$ (blue), 
	$\bar{\omega}=2.0$ (orange), $\bar{\omega}=4.0$ (green), 
	$\bar{\omega}=6.0$ (red), $\bar{\omega}=8.0$ (no strings), and $ 
	\lambda = 0.125$. Total energy for all the runs is kept fixed at, 
	$ E_{\rm Single} = 4000$,
	by adjusting $\bar{a}$ suitably according to 
	Eq.~\eqref{EnergySingle}.}
\label{LengthOmega}
\end{figure}



\begin{figure*}[t]
\includegraphics[height=0.35\textwidth,angle=0]{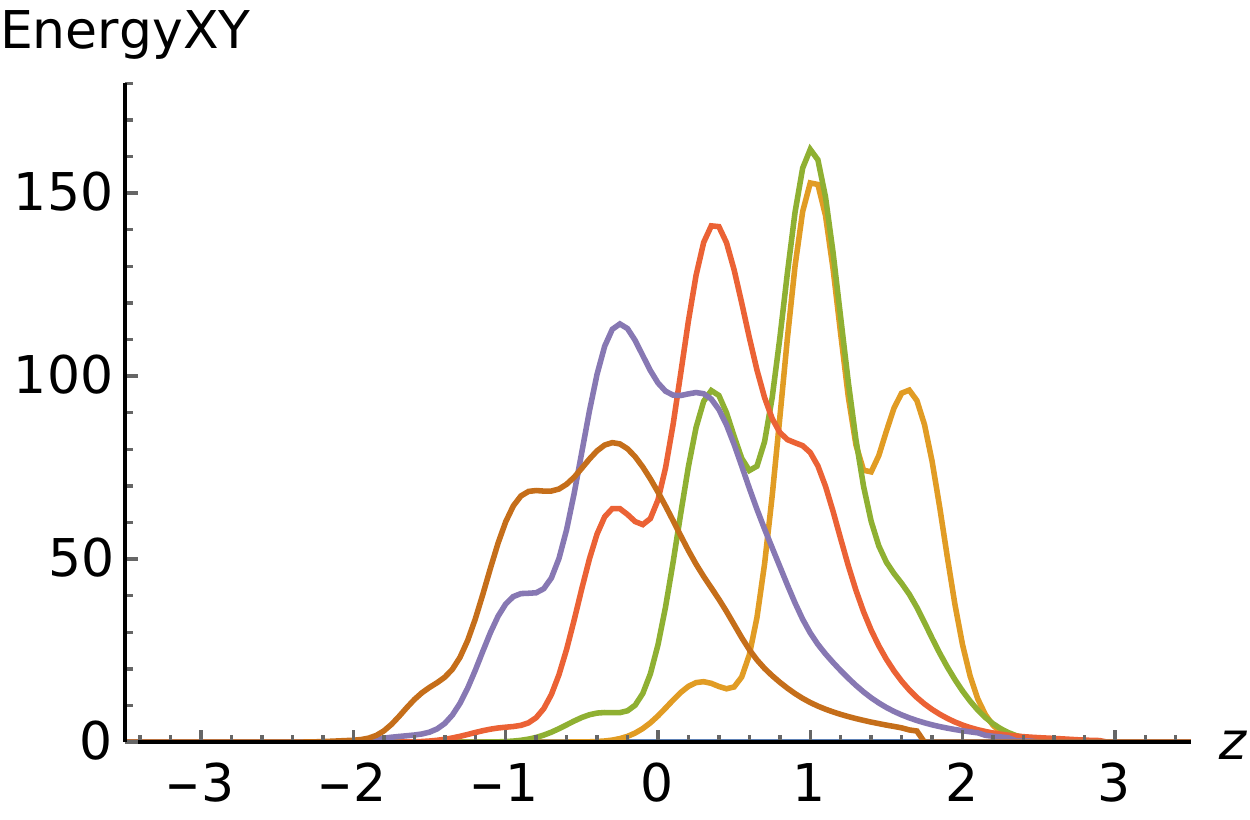}

\caption{Energy in strings in $xy-$planes as a function of $z$ at time steps $ t = 0$ (no strings), 
$t = 50$ (orange, right-most curve), 
	$ t = 100$ (green), $ t = 150$ (red), $t = 200$ (blue), and 
	$t = 250$ (brown, left-most curve),  during the simulation for $\bar{a}=6.215$, 
	$\bar{\omega}=2.0$, $ \lambda = 0.125$, and $ E_{\rm Single} = 
	4000$. Following prompt string production, the string network 
	moves to the left and decays. }
\label{SinglePulseMotion}
\end{figure*}

We have examined prompt production for several different values of the
model parameter $\lambda$ (equivalently $\beta$ since we fix $e=0.5$).
Fig.~\ref{LengthLambda} shows how the length in strings -- evaluated
by counting the plaquettes that contain non-trivial topological winding  -- 
changes with time. The figure shows that the outcome is not very sensitive
to the value of $\lambda$ and hence we set $\lambda=0.125$ ($\beta=1$)
in the runs described below. 

In contrast, as seen in Fig.~\ref{LengthOmega}, the prompt production of 
strings depends sensitively on the parameter $\bar{\omega}$. The general 
trend is
that less length is produced for larger $\bar{\omega}$ but the strings 
that are
produced survive for a longer time. This can happen if larger 
$\bar{\omega}$
leads to larger loops or to loops with higher angular momentum.

\begin{figure*}
	\begin{minipage}[t]{3in}
\includegraphics[height=0.60\textwidth,angle=0]{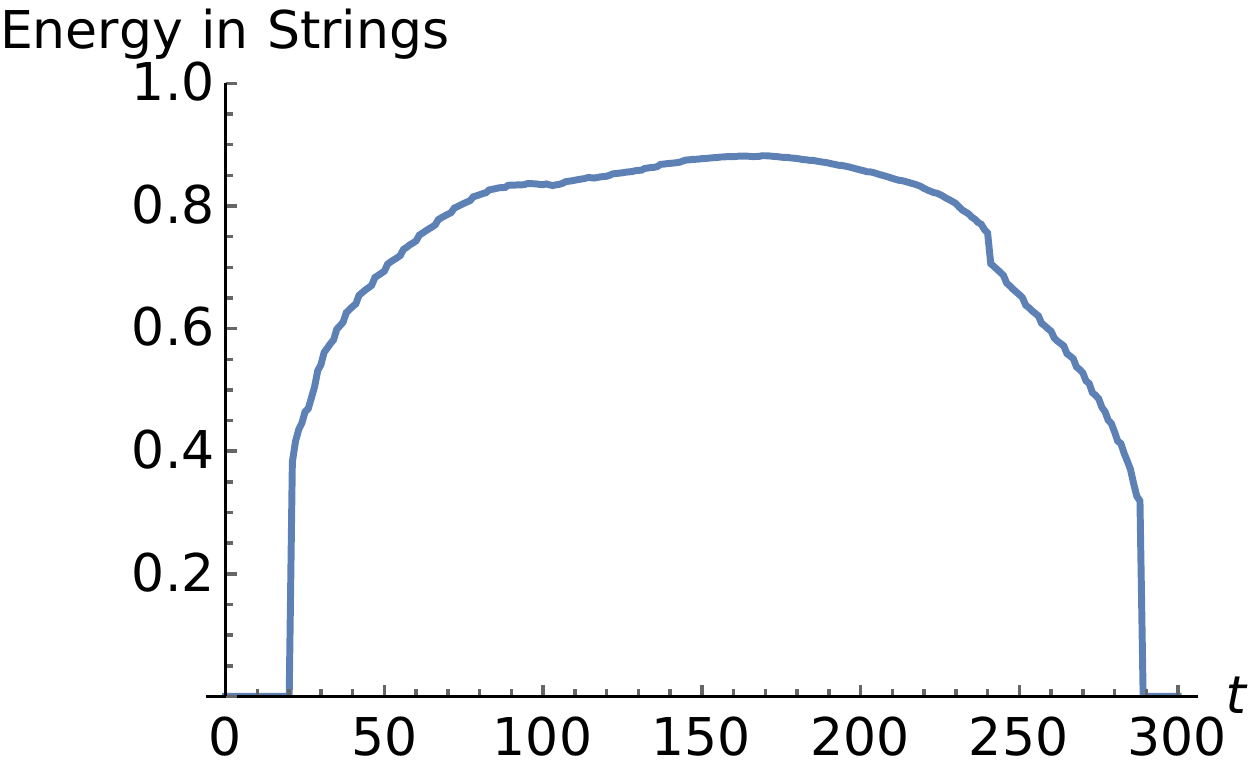}
		\caption{Energy in strings as a fraction of total energy 
		versus time(-steps) for $ \bar{a}=6.215$, 
		$\bar{\omega}=2.0$, $ \lambda = 0.125$, and $ E_{\rm 
		Single}=4000$.}
 \label{EnergyInStringsSingle}
\end{minipage}
	\hspace{1.4cm}
\begin{minipage}[t]{3.0in}
  \includegraphics[trim={0 0 0 
	0},height=0.60\textwidth,angle=0]{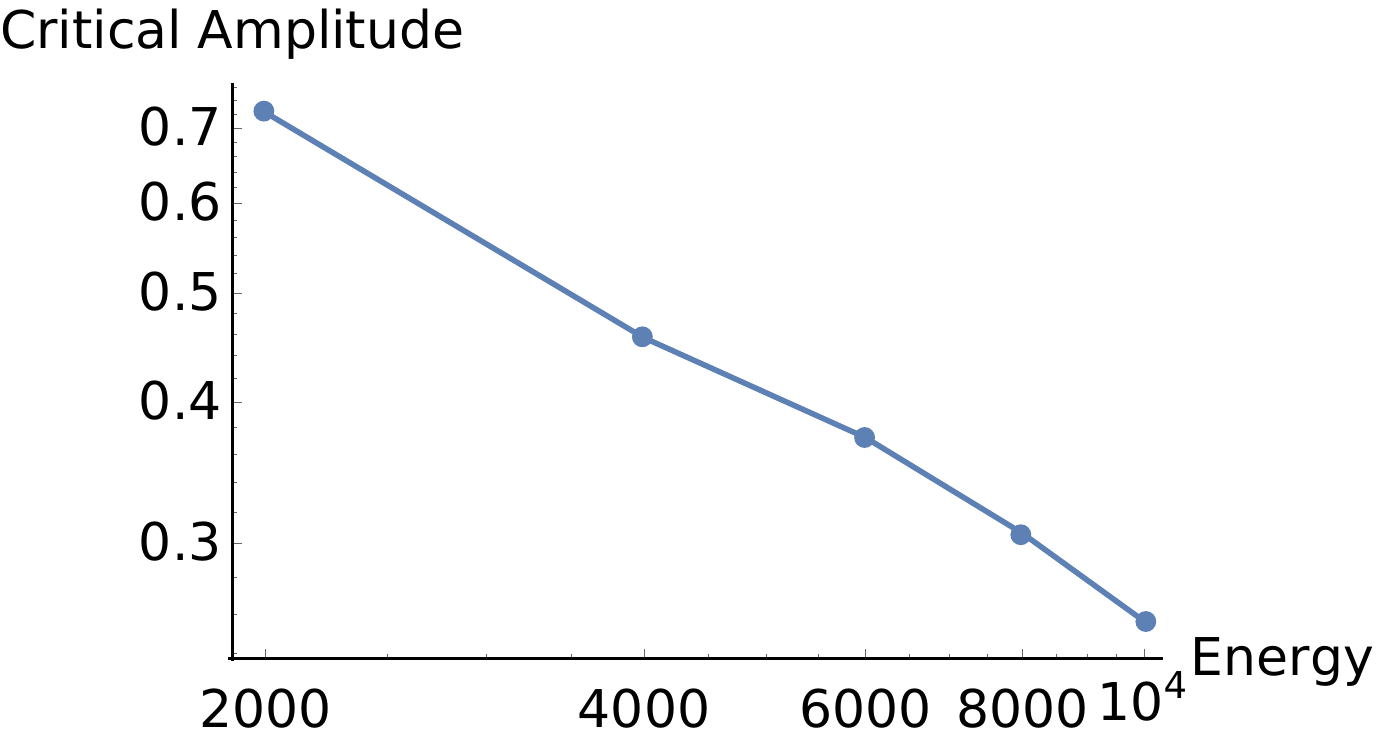}
  \caption{Critical amplitude ${\bar a}$ as we change total input energy, $E_{\rm Single}$,
  for the single pulse case with $ \lambda = 0.125$. Strings are only 
	produced above the curve.
	}
 \label{CriticalSingle}
\end{minipage}
\end{figure*}


%
In Fig.~\ref{SinglePulseMotion} we plot the energy density integrated over
$\bar{x}$ and $\bar{y}$ coordinates, as a function of $\bar{z}$.
Prompt string production occurs at the initial location of the wavepacket 
($\bar{z}_0=1.8$ or 36 lattice spacings away
from the center of the lattice). Then the string cluster moves towards the left and
also decays.

We have calculated the length of strings at any given time by counting the number of
plaquettes with non-trivial winding. We can also estimate the energy in the string network
by adding up the field energies in all the cells within $m_S^{-1}$ or $ 
m_V^{-1}$ (whichever is larger) of the string network.
However, the plot in Fig.~\ref{EnergyInStringsSingle} 
of the
energy vs. time shows reasonable correspondence with the length versus time 
plot in Fig.~\ref{LengthOmega} for $\bar{\omega}=2.0$, indicating
that the strings do not have significant kinetic energy at formation. 

As expected, greater initial energy produces more strings.  However, our 
analysis indicates some subtleties in the process of 
string production. From Eq.~(\ref{EnergySingle}), it can be seen 
that, for fixed energy, amplitude becomes smaller as we increase the 
frequency and vice-versa.  (The wavepacket width $\bar{w}$ is fixed in
all our runs.) After experimenting with different 
values of amplitude and frequency at fixed energy, we noticed that there 
is a minimum/critical amplitude 
below which we do not produce any strings, as seen in Fig.~\ref{CriticalSingle}). 
The parameter space under the critical curve, for which strings are 
not produced, gets smaller as the energy increases. In the opposite limit 
of small $\bar{\omega}$ (large amplitude), we see that the total length of 
strings is far greater (also seen in Fig.~\ref{LengthOmega}).

\subsection{Wavepacket collisions}
\label{collisions}


We now consider the case when two wavepackets collide.
The parameters are chosen so that there is no prompt string production.
However, strings are produced when the wavepackets collide. So now
we have two wavepackets in the initial conditions that are headed towards
a collision. The initial energy is,
\ba
E_{\rm Double} &=& 2 E_{\rm Single} +
\frac{\pi^{3/2} \bar{a}^2 \eta}{4 \bar{w}^7} e^{-\bar{z}_0^2 / \bar{w}^2} 
\biggl [ -18 \bar{w}^2 {\bar{z}_0}^2 + 4 {\bar{z}_0}^4 \nonumber \\ &&
 + 2 \bar{w}^8 \bar{\omega}^2 (e^2 + 2 \bar{\omega}^2) + 2 \bar{w}^6 
 (e^2+5 \bar{\omega}^2) \nonumber \\
&&
+ \bar{w}^4 ( 9-2 e^2 {\bar{z}_0}^2 -8 {\bar{z}_0}^2 \bar{\omega}^2 ) 
\cos(2 \bar{z}_0 \bar{\omega} ) \nonumber \\ &&
- 8 \bar{w}^2 \bar{z}_0 \omega (2 \bar{w}^2 - {\bar{z}_0}^2 + \bar{w}^4 
\bar{ \omega}^2 ) \sin(2 \bar{z}_0 \bar{ \omega}) \biggr ]
\ea


We again use Eq.~\eqref{EnergySingle} for fixing kinematic 
parameters.  For the simulation, we chose $\bar{a} = 0.578$ and $\bar{\omega} = 9.0$ 
for the individual wavepackets. With this choice prompt production of
strings does not occur, that is, the parameters lie below the critical curve 
for the single pulse case shown in Fig.~\ref{CriticalSingle}.
  

\begin{figure*}[h]
\includegraphics[height=0.24\textwidth,angle=0]{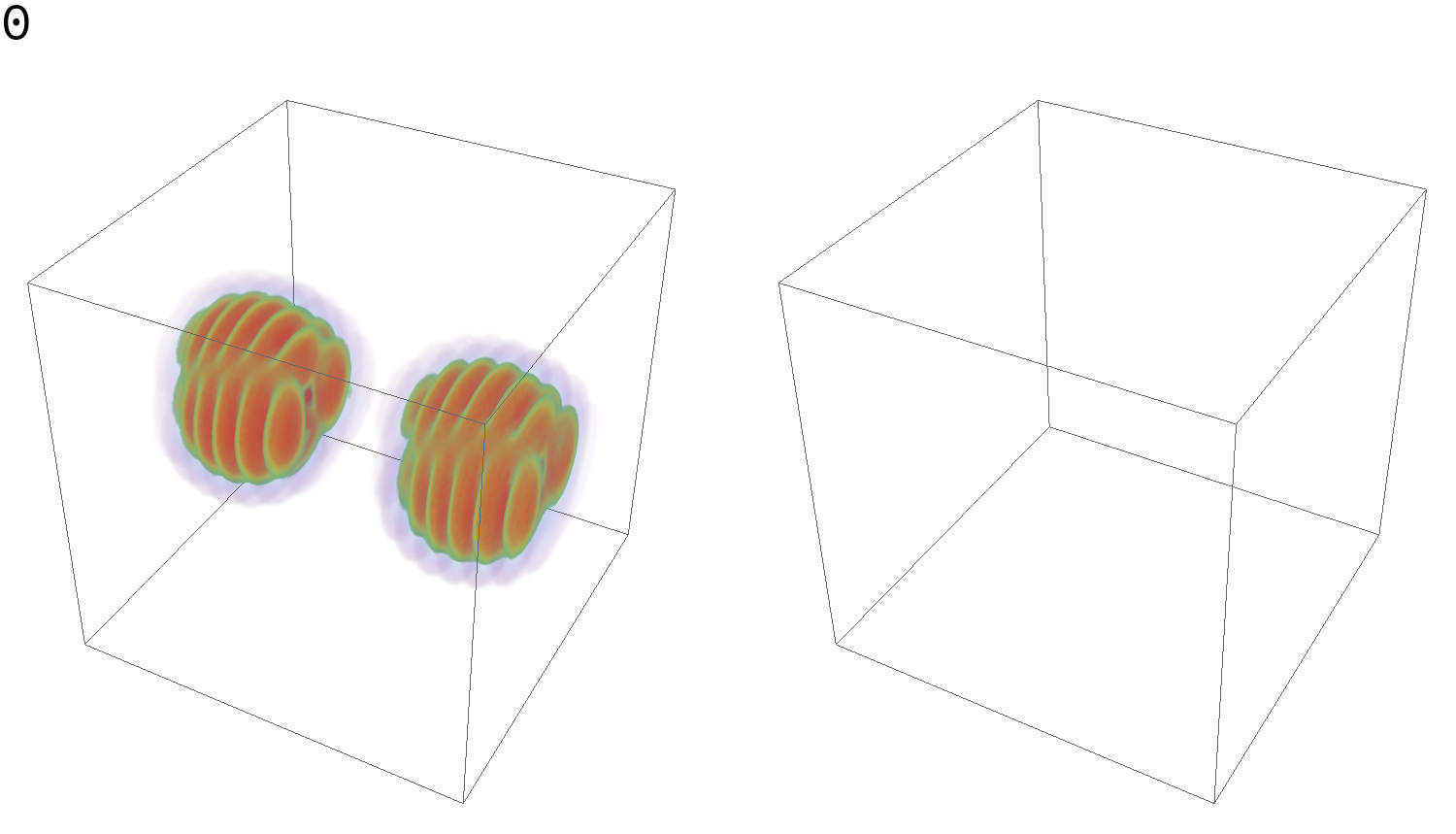}
\includegraphics[height=0.24\textwidth,angle=0]{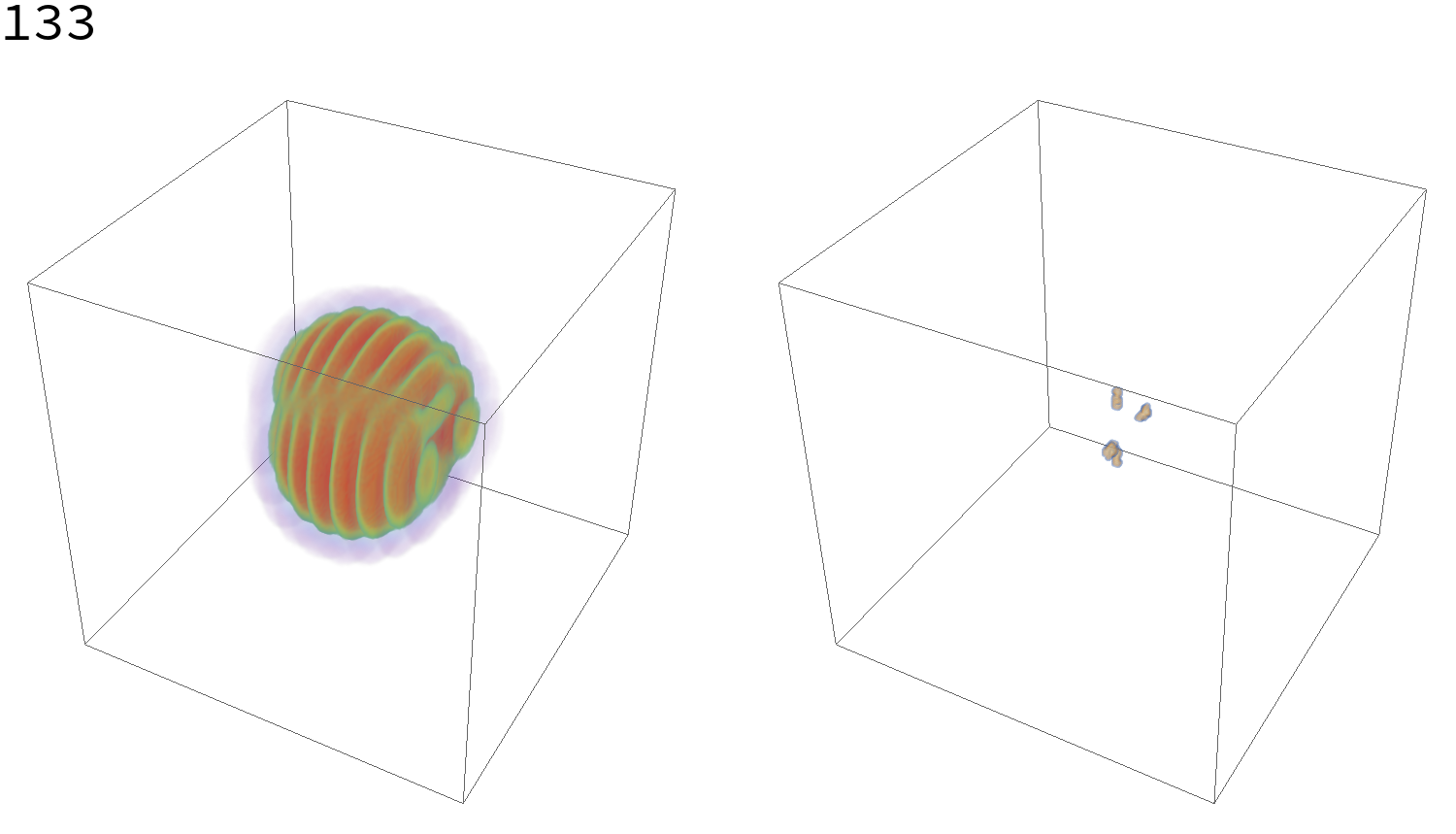}
\includegraphics[height=0.24\textwidth,angle=0]{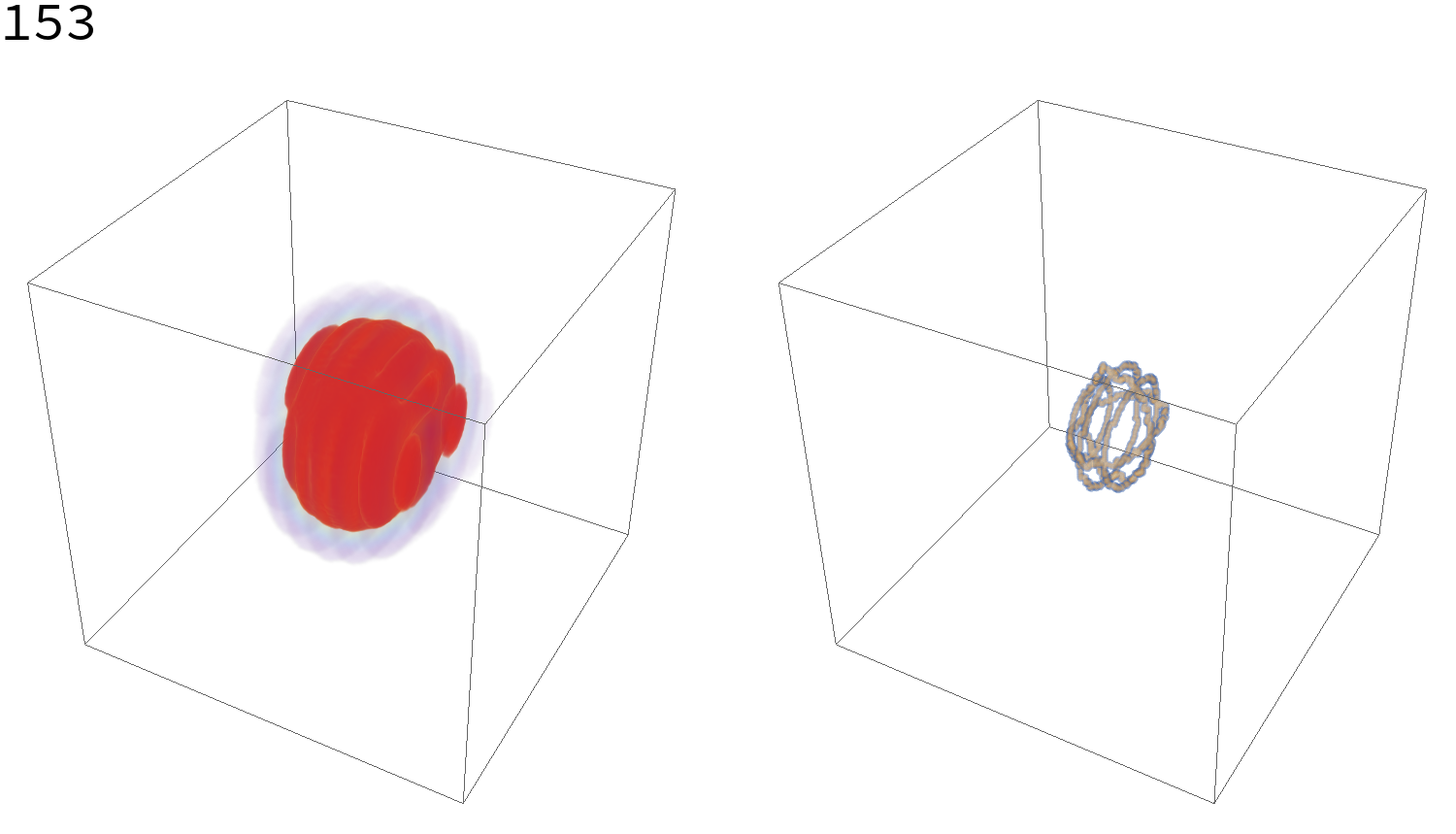}
\includegraphics[height=0.24\textwidth,angle=0]{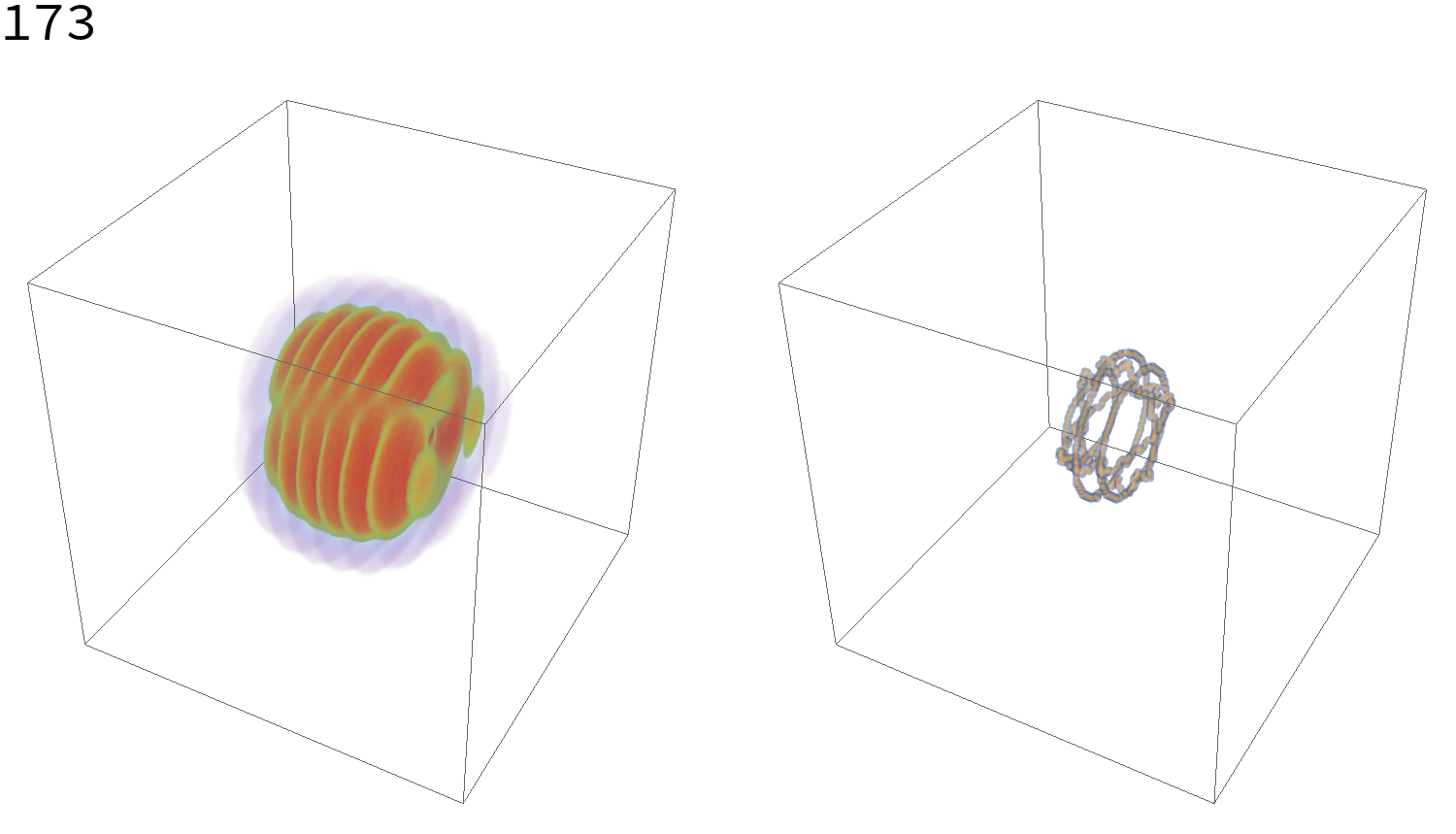}
\includegraphics[height=0.24\textwidth,angle=0]{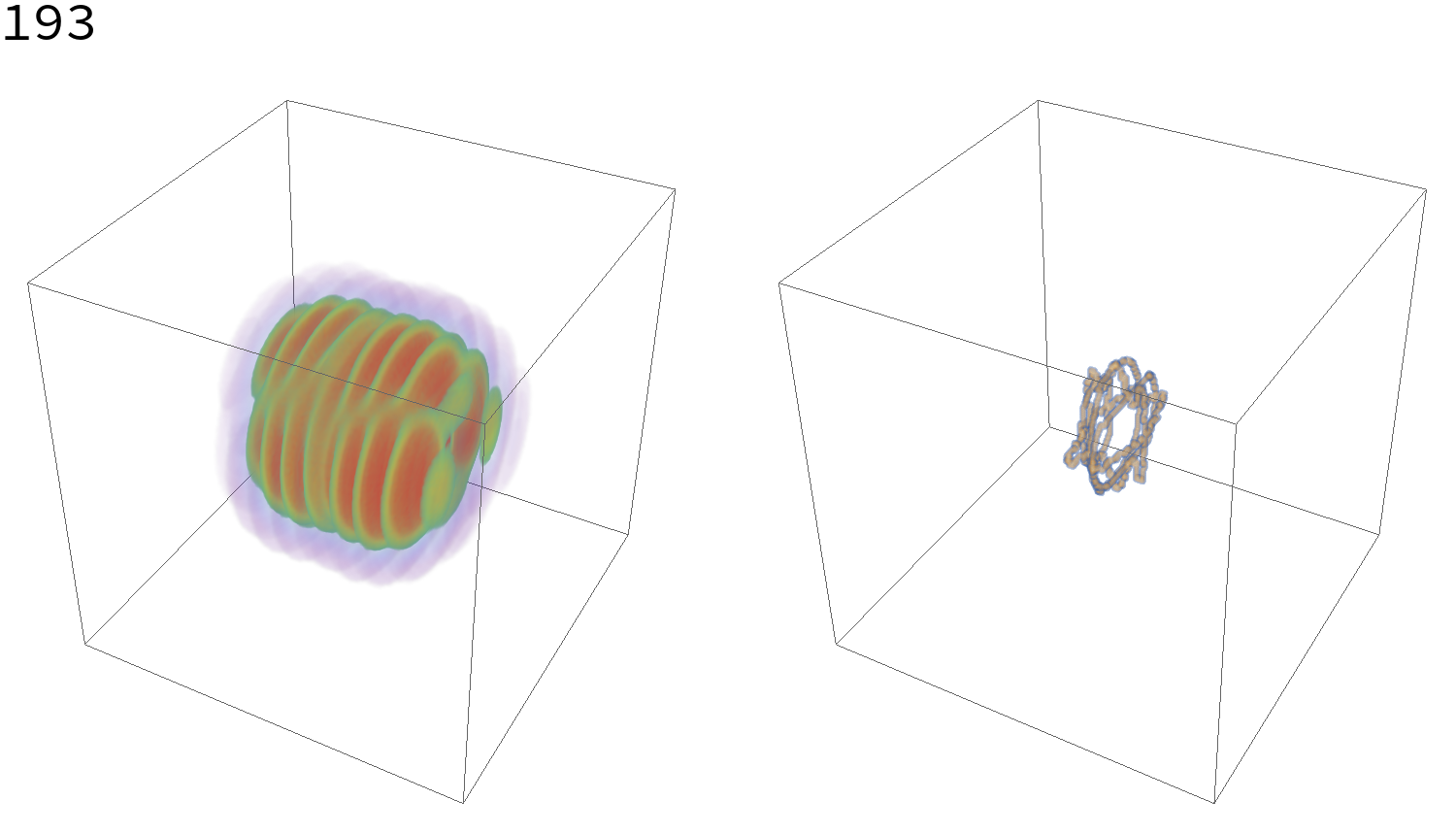}
\includegraphics[height=0.24\textwidth,angle=0]{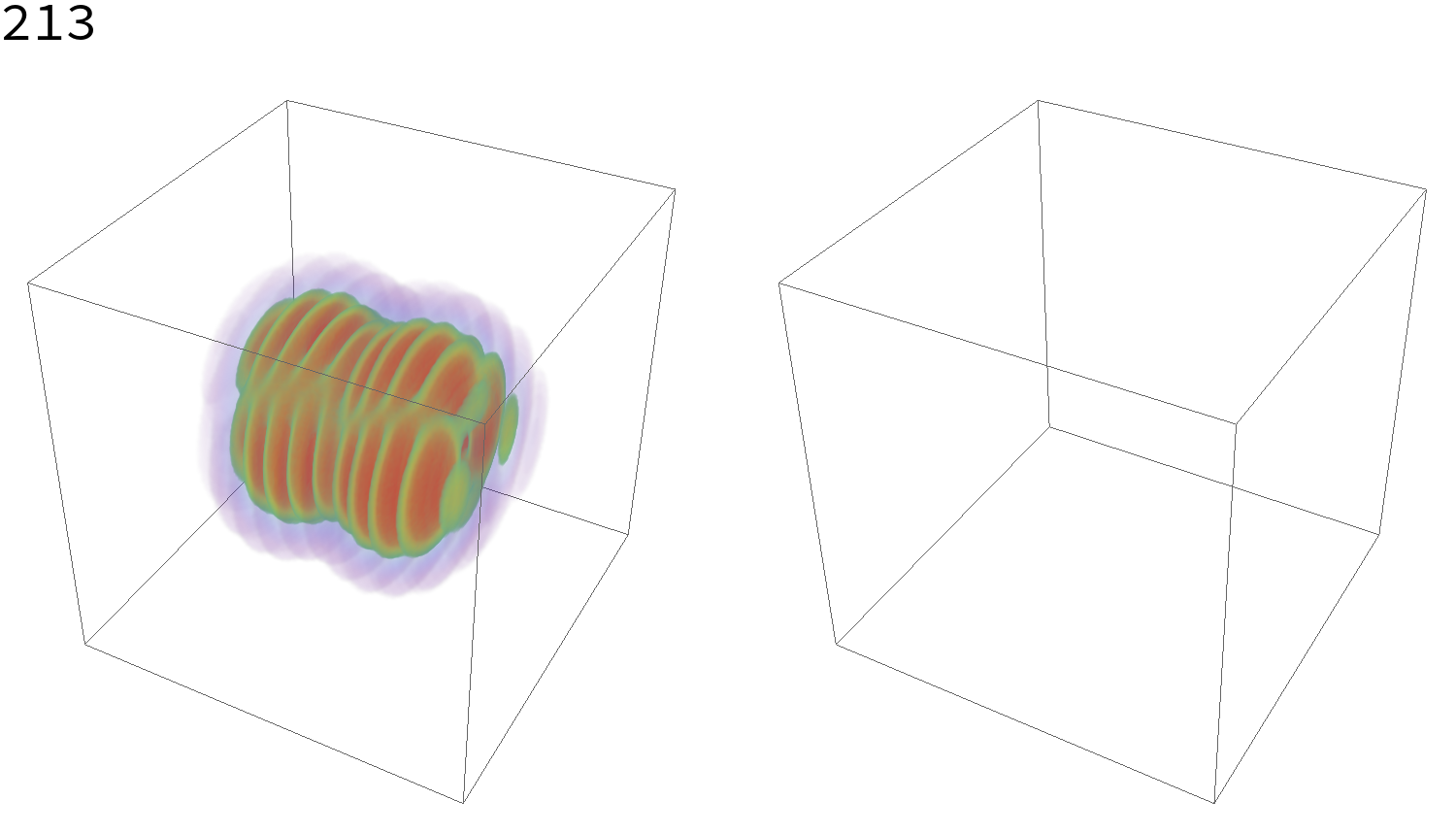}
\includegraphics[height=0.07\textwidth,angle=0]{barlegend.png}
\centering
		\caption{Total energy density (boxes on the left) and 
		winding (boxes on the right) at different time steps for 
		the case of two collinear pulses for $\bar{a}=0.578$, 
		$\bar{\omega}=9.0$, $ \lambda = 0.125$, and $E_{\rm Double} 
		\approx 8000$. The strings are first produced at time step 
		$133$ in our simulation, and therefore we have not shown 
		plots for intermediate time steps. The $116^3$ boxes shown 
		here are smaller than the full lattice ($256^3$).
		}
\label{TwoPulsesMovie}
\end{figure*}

\begin{figure*}
\begin{minipage}[t]{3.0in}
  \includegraphics[height=0.60\textwidth,angle=0]{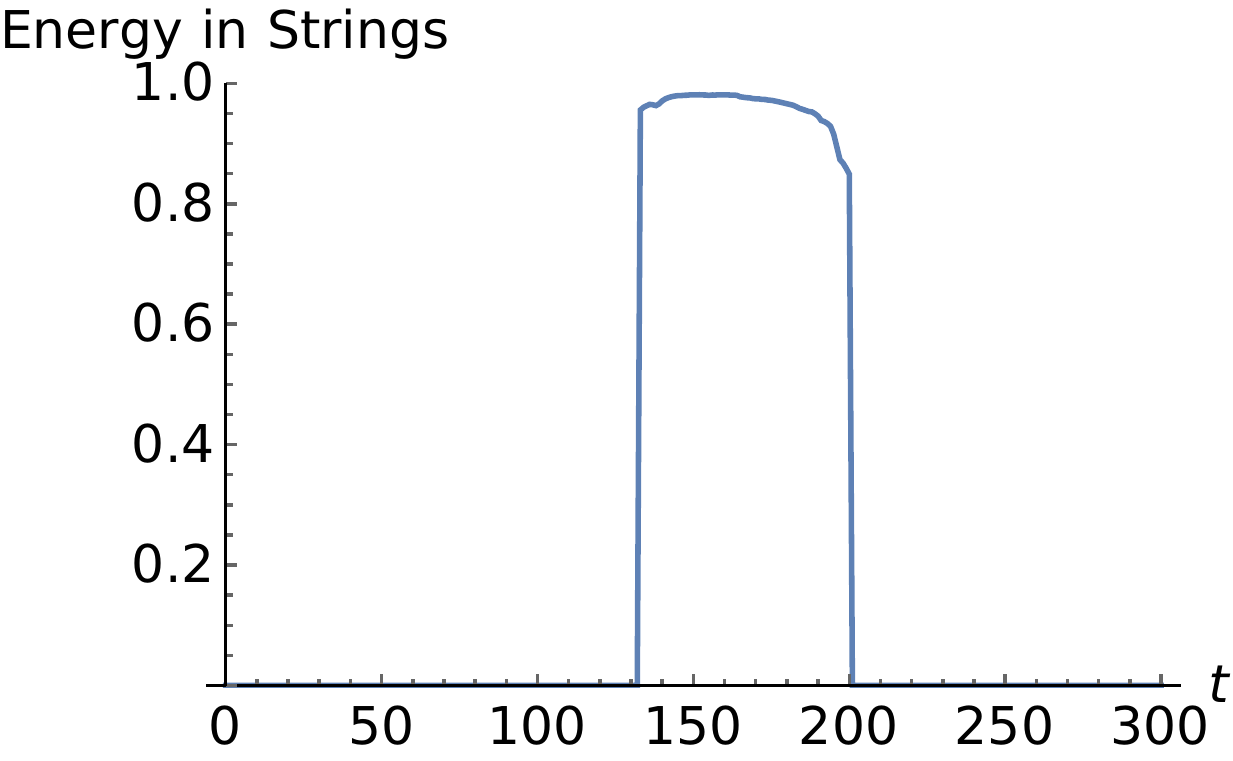}
	\caption{Energy in strings as a fraction of total energy versus 
	time for $ \bar{a}=0.578$, $\bar{\omega}=9.0$, $ \lambda = 0.125$,  
	and $ E_{\rm Double} \approx 8000$. }
\label{EnergyInStringsDouble}
\end{minipage}
\hspace{1.4cm}
\begin{minipage}[t]{3.0in}
  \includegraphics[height=0.60\textwidth,angle=0]{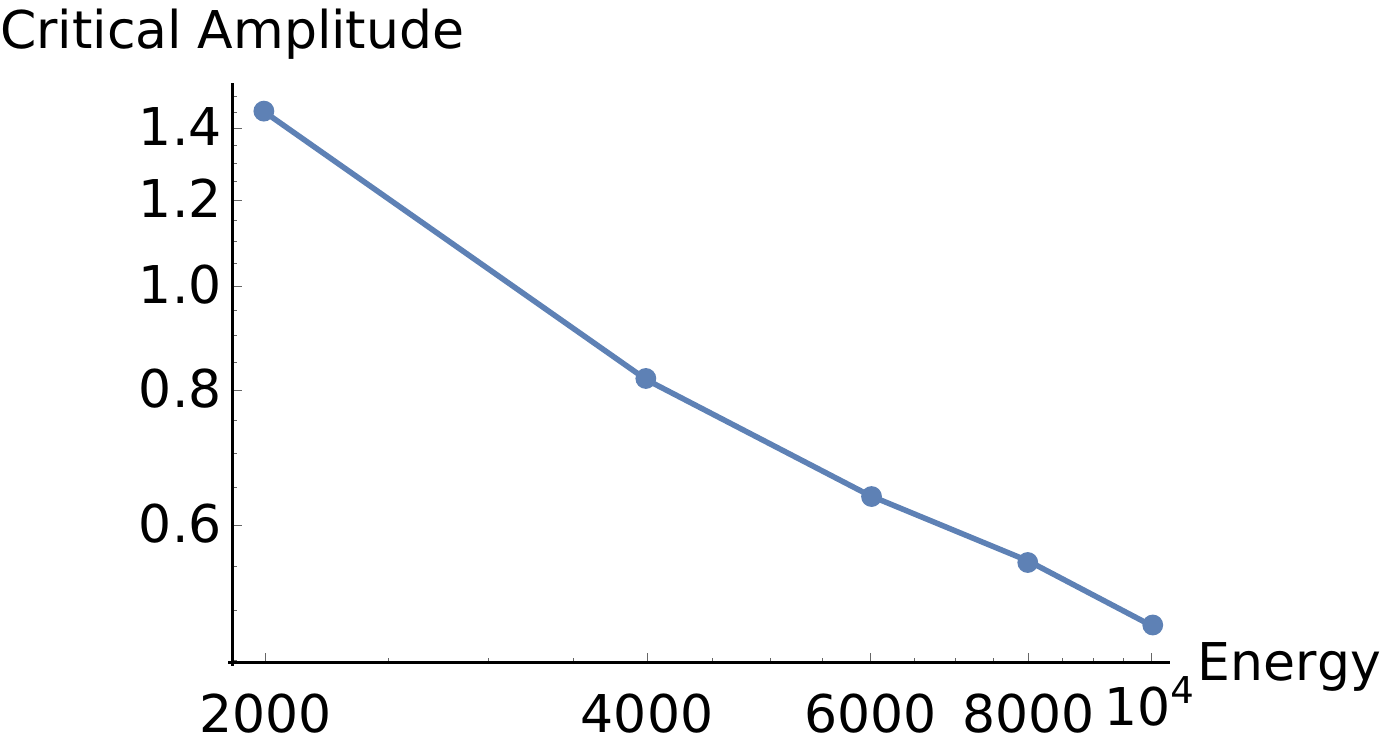}
  \caption{Critical amplitude ${\bar a}$ as we change total input energy, 
  $E_{\rm Double}$,
  for the case of colliding wavepackets for $ \lambda = 0.125$. Strings 
  are only produced above the curve.}
\label{CriticalDouble}
\end{minipage}
\end{figure*}

Fig.~\ref{TwoPulsesMovie} shows the evolution of the wavepackets and
string formation after collision. Very few short-lived strings are 
produced even though the total input energy is much higher ($\approx 
8000$) compared to the single pulse run presented in the previous 
subsection.  The fractional energy in strings as a function of time is
shown in Fig.~\ref{EnergyInStringsDouble}. By scanning over different amplitudes, 
${\bar a}$, for the same total 
energy, we find the critical curve for string formation when wavepackets 
collide. The critical
curve is plotted in Fig.~\ref{CriticalDouble}.

%
%

\section{Conclusions}
\label{conclusions}

We have explored the formation of U(1) gauge strings due to wavepackets of
of gauge fields in two settings: (i) the prompt formation of strings from gauge
fields, and (ii) the formation of strings when gauge wavepackets collide. We have
restricted our attention to a class of wavepackets with certain parameters, and 
found critical curves in parameter space that demarcate string formation regions.

It is interesting to contrast string production with magnetic monopole production.
Unlike the case of magnetic monopoles, the string loops that are formed are 
short-lived as they collapse and produce radiation. The loops may live longer
if we could find initial conditions that provide them with greater angular momentum
but these too will not live indefinitely. On the other hand, once a magnetic monopole
and antimonopole pair are produced with sufficient velocity, they will move apart and
survive indefinitely. Furthermore, magnetic monopoles are localized objects and
so the colliding wavpackets need not be very extended. For strings, the wavepackets
have to extend over a region that is the size of the string loop that is to be produced,
and only relatively small loops can be produced.
In these respects it appears that magnetic monopoles are easier to produce than
strings. 

The flip side is that we know systems that contain gauge strings while
the existence of magnetic monopoles is still speculative.
Gauge strings are known to exist in superconductors and, in that setting, our gauge field 
wavepackets correspond to photon wavepackets. This suggests that by shining light on 
superconductors we could produce strings within the superconductor. However, a realistic 
superconductor is described by a different set of equations that take into account the 
dependence of the model parameters on the temperature~\cite{Millis}. 
It will be interesting to adapt our analysis to study string production in superconductors.

\section{Acknowledgment}
We are grateful to Onur Erten for remarks.
AS thanks the MCFP, University of Maryland for hospitality. The computations were done 
on the A2C2 Saguaro and Agave clusters at ASU.
This work is supported by the U.S. Department of Energy, Office of High Energy 
Physics, under Award No.~\uppercase{DE-SC0018330} at ASU.

\newpage
\bibliography{main}

\begin{thebibliography}{6}
\expandafter\ifx\csname natexlab\endcsname\relax\def\natexlab#1{#1}\fi
\expandafter\ifx\csname bibnamefont\endcsname\relax
  \def\bibnamefont#1{#1}\fi
\expandafter\ifx\csname bibfnamefont\endcsname\relax
  \def\bibfnamefont#1{#1}\fi
\expandafter\ifx\csname citenamefont\endcsname\relax
  \def\citenamefont#1{#1}\fi
\expandafter\ifx\csname url\endcsname\relax
  \def\url#1{\texttt{#1}}\fi
\expandafter\ifx\csname urlprefix\endcsname\relax\def\urlprefix{URL }\fi
\providecommand{\bibinfo}[2]{#2}
\providecommand{\eprint}[2][]{\url{#2}}

\bibitem[{\citenamefont{Vachaspati}(2016)}]{TanmayCreation}
\bibinfo{author}{\bibfnamefont{T.}~\bibnamefont{Vachaspati}},
  \bibinfo{journal}{Phys. Rev. Lett.} \textbf{\bibinfo{volume}{117}},
  \bibinfo{pages}{181601} (\bibinfo{year}{2016}),
  \urlprefix\url{https://link.aps.org/doi/10.1103/PhysRevLett.117.181601}.

\bibitem[{\citenamefont{Baumgarte and Shapiro}(2010)}]{NumericalRelativity}
\bibinfo{author}{\bibfnamefont{T.~W.} \bibnamefont{Baumgarte}}
  \bibnamefont{and} \bibinfo{author}{\bibfnamefont{S.~L.}
  \bibnamefont{Shapiro}}, \emph{\bibinfo{title}{Numerical Relativity: Solving
  Einstein's Equations on the Computer}} (\bibinfo{publisher}{Cambridge
  University Press}, \bibinfo{address}{New York, NY, USA},
  \bibinfo{year}{2010}), ISBN \bibinfo{isbn}{052151407X, 9780521514071}.

\bibitem[{\citenamefont{Vachaspati and Vilenkin}(1984)}]{VilenkinVachaspati}
\bibinfo{author}{\bibfnamefont{T.}~\bibnamefont{Vachaspati}} \bibnamefont{and}
  \bibinfo{author}{\bibfnamefont{A.}~\bibnamefont{Vilenkin}},
  \bibinfo{journal}{Phys. Rev. D} \textbf{\bibinfo{volume}{30}},
  \bibinfo{pages}{2036} (\bibinfo{year}{1984}),
  \urlprefix\url{https://link.aps.org/doi/10.1103/PhysRevD.30.2036}.

\bibitem[{\citenamefont{Pogosian and Vachaspati}(1998)}]{TanmayLevonGeodesic}
\bibinfo{author}{\bibfnamefont{L.}~\bibnamefont{Pogosian}} \bibnamefont{and}
  \bibinfo{author}{\bibfnamefont{T.}~\bibnamefont{Vachaspati}},
  \bibinfo{journal}{Physics Letters B} \textbf{\bibinfo{volume}{423}},
  \bibinfo{pages}{45 } (\bibinfo{year}{1998}), ISSN \bibinfo{issn}{0370-2693},
  \urlprefix\url{http://www.sciencedirect.com/science/article/pii/S0370269398001099}.

\bibitem[{\citenamefont{Hindmarsh and Rajantie}(2000)}]{Hindmarsh:2000kd}
\bibinfo{author}{\bibfnamefont{M.}~\bibnamefont{Hindmarsh}} \bibnamefont{and}
  \bibinfo{author}{\bibfnamefont{A.}~\bibnamefont{Rajantie}},
  \bibinfo{journal}{Phys. Rev. Lett.} \textbf{\bibinfo{volume}{85}},
  \bibinfo{pages}{4660} (\bibinfo{year}{2000}), \eprint{cond-mat/0007361}.

\bibitem[{\citenamefont{Kennes and Millis}(2017)}]{Millis}
\bibinfo{author}{\bibfnamefont{D.~M.} \bibnamefont{Kennes}} \bibnamefont{and}
  \bibinfo{author}{\bibfnamefont{A.~J.} \bibnamefont{Millis}},
  \bibinfo{journal}{Phys. Rev. B} \textbf{\bibinfo{volume}{96}},
  \bibinfo{pages}{064507} (\bibinfo{year}{2017}),
  \urlprefix\url{https://link.aps.org/doi/10.1103/PhysRevB.96.064507}.

\end{thebibliography}

\end{document}